# Improved Buildup Model for Radiation-Induced Defects in MOSFET Isolation Oxides

[a]Hesham. H. Shaker,   [a]A.A. Saleh,   [b]Mohamed Refky Amin,   [b]S. E. D. Habib

[a] Dept. of Nuclear Safety Research and Radiological Emergencies – NCRRT Center – EAEA

[b] Dept. of Electronics and Electrical Communications –  Faculty of Eng. – Cairo University

Egypt

*Abstract*— Ionizing radiation induces defects in STI oxides in current MOSFETs. These defects may degrade the performance of the MOS circuit. Analytical models for the buildup of these defects during the radiation exposure are available in literature. In this paper, we show that the classical model used to estimate the buildup of TID-induced traps in MOSTs predicts inaccurate results at high radiation levels. We, further, introduce an improved model to estimate the buildup of defects that is valid for both low and high radiation doses. Our improved model is compared to published data showing its validity.

## I. INTRODUCTION

The gamma rays are the main threatening ionizing radiation on electronic equipment mounted in nuclear facilities [1]. One of the gamma radiation effects on ICs is the Total Ionizing Dose (TID) effects. TID effects are translated into radiation induced defects (oxide traps and interface traps) inside the oxide portions of the ICs. These defects buildup during the gamma exposure time and may cause functionality failure for the exposed ICs [2][3][4].

Radiation Hardening By Design (RHBD) approach is the most attractive approach for designing radiation hardened ICs as it doesn't need specially developed fabrication processes or heavy shields. The RHBD approach involves a number of hardening techniques during the design phase. The evaluation of the effects of the ionizing radiation on an IC before fabrication is an important part of the RHBD approach. Such evaluation can be carried out effectively by circuit simulators if the radiation-aware MOSFET models are available [1][5][6].

In 2009, Mclean [7] developed an analytical model that estimates the buildup of the TID-induced oxide traps only versus time (assuming neglected interface traps densities for low doses). In 2011, Esqueda [8][9] included the calculations of interface traps density ($D_{it}$) to the analytical defect buildup model of Mclean. Petukhov et al [10] developed a slightly different model (w.r.t Esqueda's model) to calculate the TID-induced traps concentrations up to less than 1Mrad doses. In this paper, we show that the classical model used to estimate the buildup of TID-induced traps in MOSTs predicts inaccurate results at high radiation levels. We further introduce an improved model for TID-induced defects in STI oxides of current MOSFETs that extends this model to high dose levels.

Section II illustrates the defect buildup model of Esqueda. Our improvements to this model are clarified in Section III. Then, Section IV reviews previously published experimental data that we use to validate our improved model. Section V discusses the implementation and validation of our improved model. Finally, Section VI compares the model results of our improved model and Esqueda's model.

## II. MODELING RADIATION-INDUCED DEFECTS BUILDUP IN ISOLATION OXIDES

As the ionizing radiation passes through a silicon-dioxide film, energy is transferred to the film, resulting in generation of e-h pairs (ehps). The total energy absorbed in the $SiO_2$ is called Total Ionizing Dose (TID) which is defined as the energy absorbed by the material per unit mass, and its conventional unit is "Rad". The absorbed dose is related to the number of generated ehps by the conversion factor $g_0$ which is defined as the number of ehps generated per unit dose (in rad) and unit volume (in $cm^{-3}$) and it is equal to 8.1E12 $ehps.rad^{-1}.cm^{-3}$ for $SiO_2$ film [8][7][11].

Following the generation of ehps, some of them recombine again reducing the initial density. The ratio of ehps that escape from initial recombination to the initially generated ones is called "fractional yield (Y)". Fractional yield is dependent on the applied electric field, and the type of the incident ionizing radiation. Equation 1 presents the empirical formula that relates the fractional yield to the applied electric field for gamma rays radiated from Cobalt 60 isotope ($Co^{60}$) [12].

$$Y(E)=\left(\frac{0.5}{|E|}+1\right)^{-0.7} \quad (1)$$

as |E| stands for the absolute value of the electric field in $MV.cm^{-1}$.

After the radiation induced processes of generation and recombination of ehps, four time dependent processes happen: carriers transportation; carriers trapping; carriers de-trapping; and interface traps formation. As clarified in Figure 1, assume a silicon dioxide film, manufactured over a grounded bulk silicon substrate, is positively biased. The electron mobility in silicon-dioxide material is very large compared to the hole mobility. Therefore, the radiation-generated electrons migrate easily under the effect of the applied electric field and are swept out from the positive terminal. Contrarily, holes move in the opposite direction using hopping mechanisms between shallow traps until they get trapped near Si-SiO$_2$ interface. These trapped holes can be compensated by tunneled electrons from the Si substrate, or by radiation-generated electrons passing near it [13].

During the transportation of the TID-induced holes inside the oxide, holes may react with a hydrogen containing defects (called DH centers), resulting in a release of protons (H$^+$). The released protons move under the effect of a positive electric field towards the Si-SiO$_2$ interface, where it can react with a passivated dangling bonds to form interface traps.

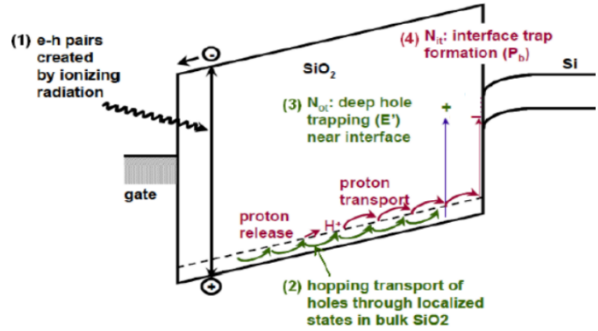

Figure 1: Band diagram of an MOS capacitor with a positive gate bias, indicating major physical processes underlying radiation response [2]

Mclean [5] developed an analytical model that estimates the buildup of the TID-induced oxide traps versus time (assuming negligible interface traps densities for low doses). Esqueda in [6] included the calculations of interface traps density ($N_{it}$) to the analytical defect buildup model of Mclean. Figure 2B depicts the charge density $\rho(x)$ distribution inside the MOSFET exposed to radiation, while Fig. 2A depicts the corresponding electric field distribution in the oxide layer. Mclean assumed the deep trapping sites are concentrated within a very small distance $x_2$ from the SiO$_2$/Si interface (called region2). The distribution of the trapped holes is assumed uniform with a density of $p_t$ per unit volume. When the radiation-induced holes are trapped at the deep trapping sites, they cause a gradual buildup of positive charges of density $qp_t(t)$ near the interface (cf. Figure 2B). These positive charges increase the localized electric field ($E_2$) at the interface. Since the density of trapped charges is assumed constant in Region 2, the electric field profile within distance $x_2$ changes linearly with an approximate slope $\frac{q}{\epsilon_{ox}} \cdot p_t$, where $\varepsilon_{ox}$ is the permittivity of the silicon dioxide. Thus, $E_1$ and $E_2$ are related by Equation 2:

$$E_1 = E_2 - \frac{q}{\epsilon_{ox}} \cdot p_t \cdot x_2 \tag{2}$$

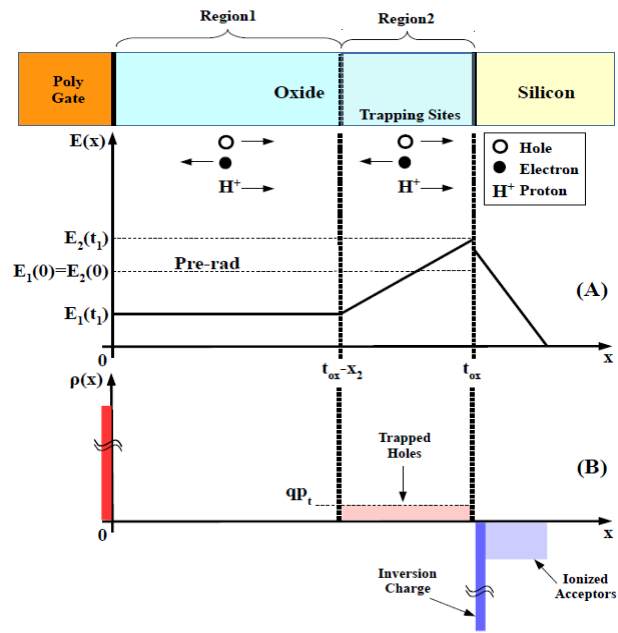

Figure 2: A) Electric field profile in oxide for pre-rad and after time radiation exposure t=t1 (for NMOS devices).
B) Charge density profile in oxide after radiation exposure time t=t1 (for NMOS devices).

The objective of the models developed in References [5] and [6] is to evaluate the evolution of $p_t(t)$ and $N_{it}(t)$ as the radiation exposure time evolves. Consider $p_t(t)$ first. The exposure time is divided into small time increments, each of width $\Delta t$. The rate equation for $p_t(t)$ {Equation 2} is invoked to calculate $\Delta p_t$ during a given time increment.

$$\frac{\Delta p_t}{\Delta t} = (N_t - p_t) \cdot \sigma_p \cdot f_p - p_t \cdot \sigma_n \cdot f_n \qquad (3)$$

Equation 3 states that the net rate of formation of oxide traps ($\Delta p_t / \Delta t$) equals to the rate of trapping TID-induced holes minus the rate of de-trapping TID-induced trapped holes. De-trapping holes is also called electron compensation, because it results from trapping TID-induced electron moving near a trapped hole. Both the trapping and de-trapping rates are zero in Region 1, which is free of trapping sites and trapped charges. $N_t$ is the density of hole trapping sites per unit volume in Region 2, $\sigma_p$ and $\sigma_n$ are the capture cross sections for holes and electrons respectively. $f_p$ and $f_n$ are the TID-induced hole and electron fluxes respectively [5]. The continuity equations for the electrons and holes are invoked to calculate $f_p$ and $f_n$ {Equations 4-5}.

$$\frac{\partial n}{\partial t} = -\frac{\partial f_n}{\partial x} + G - R \qquad (4)$$

$$\frac{\partial p}{\partial t} = -\frac{\partial f_p}{\partial x} + G - R \qquad (5)$$

$G$ is the generation rate of the radiation-induced ehps, $R$ is the recombination rate of the radiation-induced carriers. The terms  and  can be neglected, and thus, Equations 4, and 5 can be directly integrated to get $f_n$ and $f_p$. $E$ is the electric field experienced by the radiation-induced carriers, $Y$ is the fractional yield (cf. Equation 1), and $t_{ox}$ is the thickness of the dioxide film [6][5]. Therefore, we have for $E > 0$:

$$f_n = \int_x^{t_{ox}} (G - R) dx = \int_x^{t_{ox}} G Y(E) dx = D \cdot g_o \int_x^{t_{ox}} Y(E) dx \qquad (6)$$

$$f_p = \int_0^x (G - R) dx = D \cdot g_o \int_0^x Y(E) dx \qquad (7)$$

where $D$ is the gamma radiation dose rate,  is the number of ehps generated per unit dose (in rad) and unit volume and it is equal to 8.1E12 ehps.rad$^{-1}$.cm$^{-3}$ for SiO$_2$ film [6][5][7].

For case when $E < 0$, $f_n$ and $f_p$ can be written as:

$$f_n = D \cdot g_o \int_0^x Y(E) dx \qquad (8)$$

$$f_p = D \cdot g_o \int_x^{t_{ox}} Y(E) dx \qquad (9)$$

Now, we can substitute Equations 6-9 into Equation 3 to get . Since $x_2$ is small, $p_t$ is assumed constant, and to keep the model tractable, the integrals in Equations 6-9 are replaced by approximate averages. Esqueda [6] took under consideration the generated holes in Region 1 only when $E_1 > 0$, and the generated electrons in Region 1 only when $E_1 < 0$. Then, Esqueda reduced Equations 6-9 as shown in Equations 10 and 11. $Y(E)$ is the average value for fractional yield of holes and electrons, and he replaced $t_{ox} - x_2$ by $t_{ox}$ as an approximation.

For $E_1 > 0$

$$\frac{\Delta p_t}{\Delta t} = D \cdot g_o \cdot \left[ (N_{ts} - p_t) \cdot \sigma_p \cdot Y(E_1) \cdot t_{ox} - p_t \cdot \sigma_n \cdot Y\left(\frac{E_1 + E_2}{2}\right) \cdot x_2 \right] \qquad (10)$$

For $E_1 < 0$

$$\frac{\Delta p_t}{\Delta t} = D \cdot g_o \cdot \left[ (N_{ts} - p_t) \cdot \sigma_p \cdot Y\left(\frac{E_1 + E_2}{2}\right) \cdot x_2 - p_t \cdot \sigma_n \cdot Y(E_1) \cdot t_{ox} \right] \qquad (11)$$

Equations 2, 10 and 11 express $\Delta p_t$ as function of the electric field $E_2$ at the semiconductor surface (cf. Figure 2) and the total TID-trapped charge in $qp_t \cdot x_2$. The classical electrostatic treatment of MOSFET device [5,13] is next invoked to express $E_2$ as function of semiconductor surface potential $\psi_s$, and finally $\psi_s$ as function of the gate potential.

Finally, the equivalent increment in trap density at semiconductor/insulator interface $\Delta N_{ot}$ is calculated by weighting $\Delta p_t$ as shown in Equation 12.

$$\Delta N_{ot} = \int_{t_{ox}-x_2}^{t_{ox}} \frac{x}{t_{ox}} \Delta p_t(x) dx = \Delta p_t x_2 \left(1 - \frac{x_2}{2 t_{ox}}\right) \tag{12}$$

A parallel treatment is included in Ref. [6] to model the buildup of the interface traps, ending up with Equations 13 and 14. $N_{it}$ is the number of TID-induced interface traps per unit area, $N_{DH}$ is the concentration of the DH centers per unit volume, $\sigma_{DH}$ is the capture cross section of radiation-induced holes by the DH centers, $N_{SiH}$ is the number of passivated dangling bonds per unit area, $\sigma_{it}$ is the capture cross section of protons by the passivated dangling bonds. When $E_1$ is positive, the released protons in regions 1, and 2 are directed towards the interface where they may form interface traps. When $E_1$ become negative, no protons are directed towards the interface except those ones released in region2, because the electric field in this subsection is still positive. Esqueda approximated Equations 13 and 14 similar to approximations of Equations 10-11.

for $E_1 > 0$

$$\frac{\Delta N_{it}}{\Delta t} = \left(\frac{1}{2}\right) \cdot (N_{SiH} - N_{it}) \cdot \sigma_{it} \cdot N_{DH} \sigma_{DH} \cdot D \cdot g_o \cdot Y(E_1) \cdot (t_{ox})^2 \tag{13}$$

for $E_1 < 0$

$$\frac{\Delta N_{it}}{\Delta t} = \left(\frac{1}{2}\right) \cdot (N_{SiH} - N_{it}) \cdot \sigma_{it} \cdot N_{DH} \sigma_{DH} \cdot D \cdot g_o \cdot Y\left(\frac{E_1 + E_2}{2}\right) \cdot (x_2)^2 \tag{14}$$

In summary, the models of References [5] and [6] of build-up of TID-induced traps in MOSFET oxides proceed as follows:

1. Start at $t = 0$, $p_t = 0$, $N_{it} = 0$.
2. Increment exposure time by $\Delta t$.
3. Calculate $\Delta p_t$, update $p_t$, and calculate $\Delta N_{ot}$.
4. Calculate $\Delta N_{it}$, and update $N_{it}$
5. Calculate the new electric field profile in oxide ($E_1$ and $E_2$)
6. Go to step 2 if exposure time is not over.

### III. PROPOSED MODIFICATIONS

In [6], Esqueda performed some approximations. These approximations are valid for very thick oxide films that are exposed to low total ionizing doses. Our objective in this paper is to develop a more generalized model of the defects' rate equations that is suitable to defect density estimation for thick and narrow oxide thicknesses and also for low and high ionizing doses.

At high ionizing doses, the total trapped charge in Region 2, $q \cdot p_t \cdot x_2$, is so large that $E_1$ becomes negative. Figure 3 depicts the electric field distribution in this case. If $E_1$ becomes comparable to $E_2$ the average electric field in Region 2 (cf. by Equations 10, 11, 14) drops to near zero, which means very small TID-induced ehps can escape the initial recombination process in region 2. Obviously, this is not a good approximation for $f_n$ and $f_p$ in this case.

To approximate $f_n$ and $f_p$ accurately, Region 2 is subdivided, as shown in Figure 3, into two Regions: Region b where the electric field is negative and Region a where the electric field is positive. The width of Region b is $x_3$, where the intersection between Regions a and b is at the zero electric field point. The width of region a is $x_2 - x_3$. Note that for low radiation doses, $E_1 > 0$, Region b vanishes, and Figure 3 falls back to Figure 2. In this case, the corresponding electric field in it can be approximated to the average $(E_1 + E_2)/2$. In contrast, when $E_1$ becomes negative, the electric field in subdivisions 'b' and 'a' can be approximated to $E_1/2$ are $E_2/2$ respectively. As $E_1$ is going more negative, $x_3$ increases, this continues till $E_1$ saturates. We calculate the TID-generated rates of both electrons and holes in the three regions 1, a, and b separately, and sum their effect as shown below. Thus, our model equations are given by:

For $E_1 > 0$

$$\frac{\Delta p_t}{\Delta t} = D \cdot g_o \cdot \left\{ (N_{ts} - P_t) \cdot \sigma_p \cdot \left[ Y(E_1) \cdot (t_{ox} - x_2) + Y\left(\frac{E_1 + E_2}{2}\right) \cdot x_2 \right] - P_t \cdot \sigma_n \cdot \left[ Y\left(\frac{E_1 + E_2}{2}\right) \cdot x_2 \right] \right\} \tag{15}$$

$$\frac{\Delta N_{it}}{\Delta t} = \frac{1}{2} (N_{SiH} - N_{it}) \sigma_{it} N_{DH} \sigma_{DH} D \cdot g_o \cdot \left[ Y(E_1) \cdot (t_{ox} - x_2)^2 + Y\left(\frac{E_1 + E_2}{2}\right) \cdot x_2^2 \right] \tag{16}$$

For $E_1 < 0$

$$\frac{\Delta p_t}{\Delta t} = D \cdot g_o \cdot \left\{ (N_{ts} - P_t) \sigma_p \cdot \left[ Y\left(\frac{E_2}{2}\right)(x_2 - x_3) + Y\left(\frac{E_1}{2}\right) x_3 \right] - P_t \cdot \sigma_n \cdot \left[ Y(E_1)(t_{ox} - x_2) + Y\left(\frac{E_2}{2}\right)(x_2 - x_3) + Y\left(\frac{E_1}{2}\right) x_3 \right] \right\} \quad (17)$$

$$\frac{\Delta N_{it}}{\Delta t} = \frac{1}{2}(N_{SiH} - N_{it}) \sigma_{it} N_{DH} \sigma_{DH} D \cdot g_o \cdot \left[ Y\left(\frac{E_2}{2}\right)(x_2 - x_3)^2 \right] \quad (18)$$

Equations 15, and 17 show that the total trapped holes flux has three components:

i) Trapped holes due to hole flux in Region 1 $[Y(E_1)(t_{ox}-x_2)]$. If $E_1<0$, this component is nulled as the holes generated in Region 1 in this case flow towards gate.

ii) Trapped holes due to hole flux in Region b $[Y(E_1/2).x_3]$. This flux is set to zero if $E_1 > 0$ as $x_3$ equals zero in this case. If $E_1<0$, average electric field in Region b = $E_1/2$.

iii) Trapped holes due to hole flux in Region a $[Y(E_2/2).(x_2..x_3)]$. If $E_1 > 0$, as $x_3=0$, and average electric field in Region a = $(E_1+ E_2)/2$. If $E_1 < 0$, as the width of Region a = $x_2-x_3$, and average electric field in Region a = $E_2/2$.

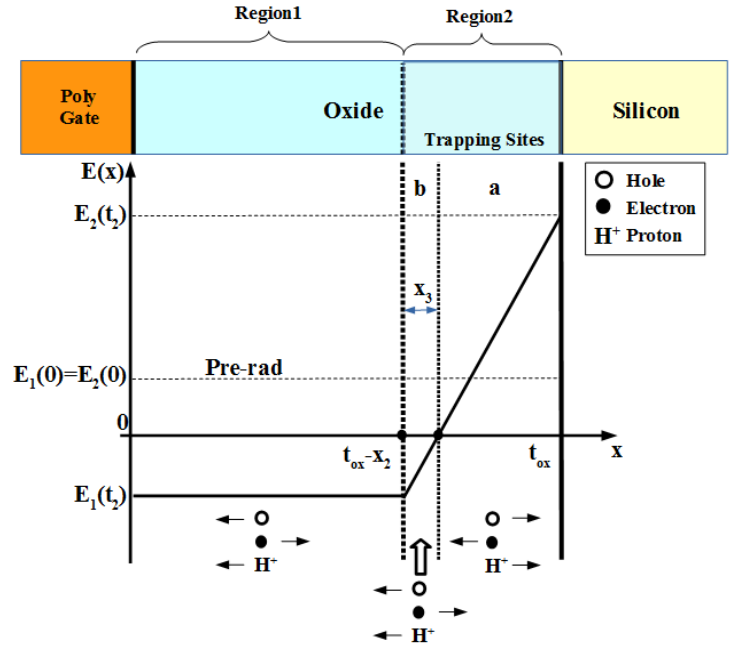

Figure 3: Electric field profile in oxide for pre-rad and after high radiation dose (for NMOS devices).

Similar explanations can be stated for de-trapping part given in Equations 15, and 17.

For the interface traps, Equations 16, and 18 calculate the rate of formation of interface traps. When $E_1$ is positive, the released protons in Regions 1, and 2 are directed towards the interface. When $E_1$ become negative, no protons are directed towards the interface except those ones released in Region 'a', because the electric field in this region is still positive. A comparison between our generalized and Esqueda's reduced models of the rate equations is presented in Table 1.

Table 1: Comparison between the formation rate equations in the Esqueda's reduced model and in the our generalized model

| Objective & Condition | Esqueda's Reduced Model Equations 10 – 17 | Our Generalized Model Equations 18 – 31 |
|---|---|---|
| Calculation of Holes Flux $E_1 > 0$ | – Calculated by E1 and used for regions 1, and 2 to get the total flux that may be trapped in region2 | – Calculated in regions 1 and 2 separately and summed to get the total flux that may be trapped in region2 |
| Calculation of Holes Flux $E_1 < 0$ | – Calculated by the average field in region 2 | |
| Calculation of Electrons Flux $E_1 > 0$ | – Calculated by the average field in region 2 | |
| Calculation of Electrons Flux $E_1 < 0$ | – Calculated by E1 and used for regions 1, and 2 to get the total flux that may be trapped in region2 | – Calculated in regions 1and 2 separately and summed to get the total flux that may be trapped in region2 |

| Calculation of Protons Flux $E_1 > 0$ | – Calculated by E1 and used for regions 1, and 2 to get the total flux that may be trapped in region2 | – Calculated in regions 1 and 2 separately and summed to get the total flux that may be trapped at the interface. |
|---|---|---|
| Calculation of Protons Flux $E_1 < 0$ | – Calculated by the average field in region 2 and used over region 2 | – Calculated for the portion of region2 which has positive electric field only. |

## IV. PUBLISHED EXPERIMENTAL RESULTS

Total ionizing dose experiments on N-well FOXFET devices, were performed by Esqueda in [8]. The structure of the N-well FOXFET device is clarified in Figure 4, the terminals of the FOXFET are like those of NMOS devices, but its drain and source contacts are attached to two N-wells, and the gate terminal is a poly-silicon which is applied over a shallow trench isolation oxide. FOXFET devices were exposed to gamma ionizing radiation with approximate dose rate 20 rad/s. During experiments the gate terminal was biased to 1V with all other terminals grounded. As plotted in Figure 5, trapped holes densities and interface traps densities were extracted at different dose levels: 20krad; 100krad; 200krad; and 1Mrad [8].

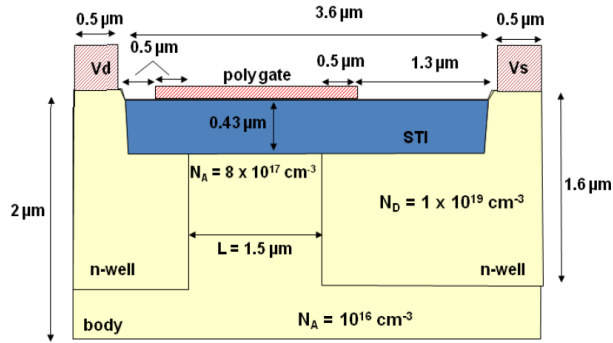

Figure 4: Cross-section of the NW FOXFET used by Esqueda in his experiments (W = 200μm, L= 1.5μm) [8].

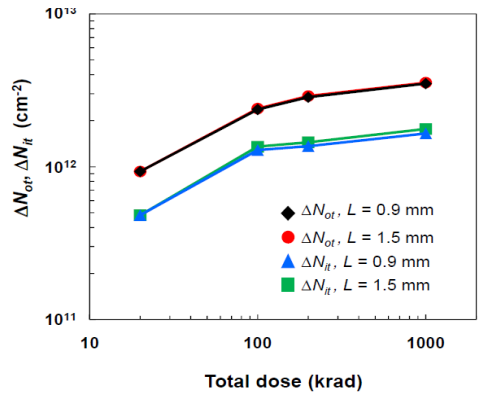

Figure 5: The extracted Not and Nit versus total ionizing dose level for two FOXFET devices of 0.9um, and 1.5um channel lengths [8].

## V. MODEL IMPLEMENTATION AND VERIFICATION

a) Solution of Surface Potential Equation (SPE):

Reference [14] presents a list of Scilab software [15] codes for various numerical methods. We selected the Bisection method to solve the Surface Potential Equation (SPE) using Scilab. The inputs of this method are: function definition, searching range of the unknown variable, maximum error between iterations, and maximum number of iterations. The outputs are the roots of the given function within the given range.

Figure 6 plots the Esqueda's numerical solution of the surface potential equation (SPE). The densities of oxide trapped charges and interface traps extracted experimentally (plotted in Figure 5), were used by Esqueda to solve the SPE at different dose levels. Figures 7 and 8 plot the solution of SPE obtained from the implemented bisection method into Scilab. As can be noticed, the solution obtained from the implemented Scilab code, matches well the Esqueda's solution.

b) Calculations of radiation induced defects buildup:

The analytical model of the defects buildup was implemented into a Scilab script. Both the Esqueda's reduced model and our generalized model of the rate equations are included in this script. The script is simulated using the same conditions of the Esqueda's experimental work in [8]. The model parameters are tweaked to get fitting between the analytical model and the experimental results. The model parameters of the Esqueda's reduced and our generalized models are tweaked separately. The model parameters used for the Esqueda's reduced and our generalized models are listed into Table 2 and Figure 9 plots the densities of defects for the simulation results and for the Esqueda's experimental work. For the oxide traps density NOT, Our generalized model fit easily the experimental results, while Esqueda's reduced model no way to fit for doses larger than 50krad. For the interface traps density NIT, our generalized and the Esqueda's reduced models can both fit the experimental results along low and high doses.

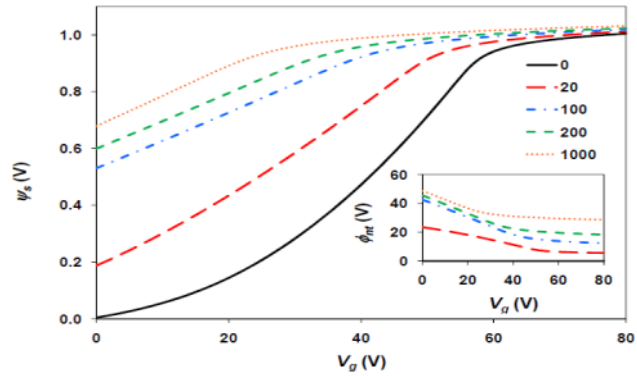

Figure 6: Surface potential versus gate voltage at different levels of total ionizing dose, and the inset shows the defect potential versus gate voltage for the same dose levels [8].

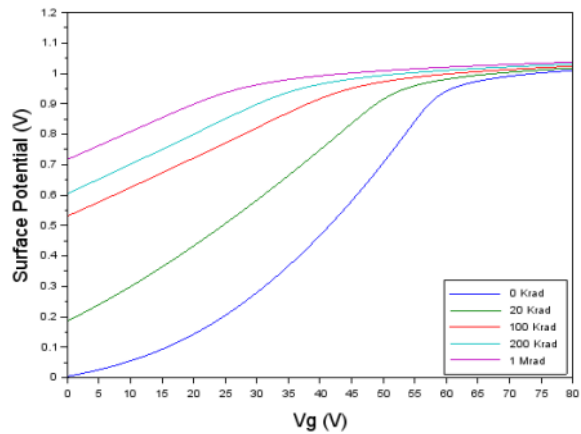

Figure 7: Our solution of the surface potential equation versus gate voltage at different levels of total ionizing dose (using Scilab).

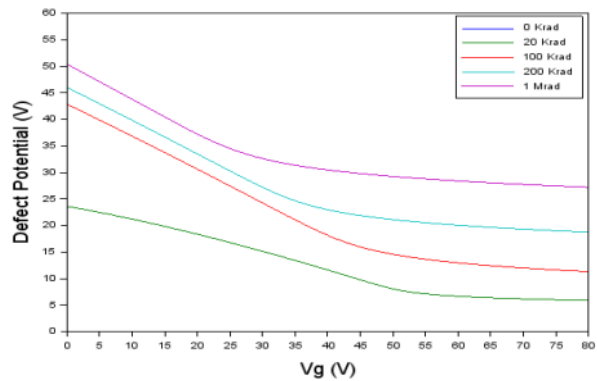

Figure 8: Calculated defect potential ($\varphi_{nt}$) versus gate voltage at different levels of total ionizing dose (using Scilab).

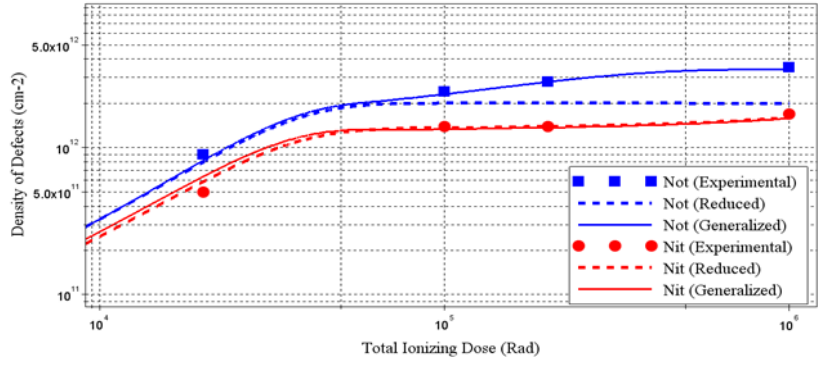

Figure 9: Densities of radiation-induced defects versus total absorbed dose

Table 2: Analytical model parameters used into the implemented Script of the defects buildup model to get the results presented in Figure 9.

| Parameter | Esqueda's Reduced Model | Our Generalized Model | Units |
|---|---|---|---|
| $N_{ts}$ | 8.0E19 | 8.0E19 | $cm^{-3}$ |
| $\sigma_p$ | 4.7E-15 | 4.5E-15 | $cm^2$ |
| $\sigma_n$ | 20E-15 | 20E-15 | $cm^2$ |
| $N_{DH}$ | 7.3E17 | 4.5E17 | $cm^{-3}$ |
| $\sigma_{DH}$ | 3E-15 | 3E-15 | $cm^2$ |
| $N_{SiH}$ | 4.8E12 | 4.8E12 | $cm^{-2}$ |
| $\sigma_{it}$ | 3.2E-12 | 3.8E-12 | $cm^2$ |
| $x_2$ | 25 | 25 | nm |

## VI. MODEL RESULTS

In this section, we compare the results of our generalized model and Esqueda's reduced model for an NMOS capacitor of an oxide thickness 425nm, substrate doping 7.4e17 $cm^{-3}$, and bias of 1 volt. We simulated both models with the same parameters values listed in the second column of Table 2. The comparison is carried out for low and high radiation doses. Figures 10 – 13 show respectively the variation of the trapping/de-trapping rates, the net trapping rate, the electric field and the densities of traps/ surface states with the TID dose up to 1E8 rad

As clarified in Figures 10, 11, 12, and 13 for the Esqueda's reduced model, when $E_1>0$, the flux of the radiation-generated holes is calculated with respect to the electric field $E_1$ along all the oxide thickness, while our generalized model calculates this flux in regions 1 and 2 separately and evaluates the net flux by adding the two calculated fluxes to each other. As the total absorbed dose in an oxide increases, more holes are trapped in Region 2, $E_2$ increases, $E_1$ decreases, hole trapping rate decreases, and holes de-trapping (electrons trapping by the trapped holes) rate increases, this process continues till the trapping and de-trapping rates are approximately equal causing a saturation in the trapped holes density. In Esqueda's model, this saturation always happens before the inversion of $E_1$ and no more changes appear in $E_1$ and $E_2$ fields. In contrast, the trapped hole density in our generalized model can saturate after the inversion of $E_1$ due to the added holes flux in region 2 before the inversion of $E_1$.

Next, we discuss radiation-induced interface states at the oxide/silicon interface. When $E_1$ is greater than zero, all the moving radiation-generated holes, in Regions 1 and 2, may interact with hydrogenated defects in the oxide film, resulting in a release of positive protons that are swept under the effect of the electric field towards the interface, where they can de-passivate dangling bonds there to form interface states. When $E_1$ becomes negative at high total doses, Eaqueda assumes positive electric field in the whole of Region 2. So he takes the contribution of interface states formation from any generated protons in Region 2. While our generalized model divides Region 2 into two subsections. So, we take the contribution of interface states formation from only one subsection in Region 2 (subsection at the side of the interface) which still has a positive field. As the total dose increases, surface potential at the interface increases. When the surface potential exceeds the bulk potential, the radiation induced interface states are get negatively charged. Then, when the density of interface states become comparable to the density of oxide traps, the defect potential ($\varphi_{nt}$) decreases, causing a decrease in $E_2$, which then decreases the density of the oxide traps. This degradation continues until the saturation of the density of interface states. The saturation of the density of interface traps happens because all the dangling bonds at the interface are de-passivated. These circumstances during interface states formation explain the occurrence of a hump in the oxide traps density around 1Mrad total dose level. This hump doesn't appear in results of Esqueda's model in Figures 12 and 13 because the trapped holes density always saturates before the inversion of $E_1$ in his model.

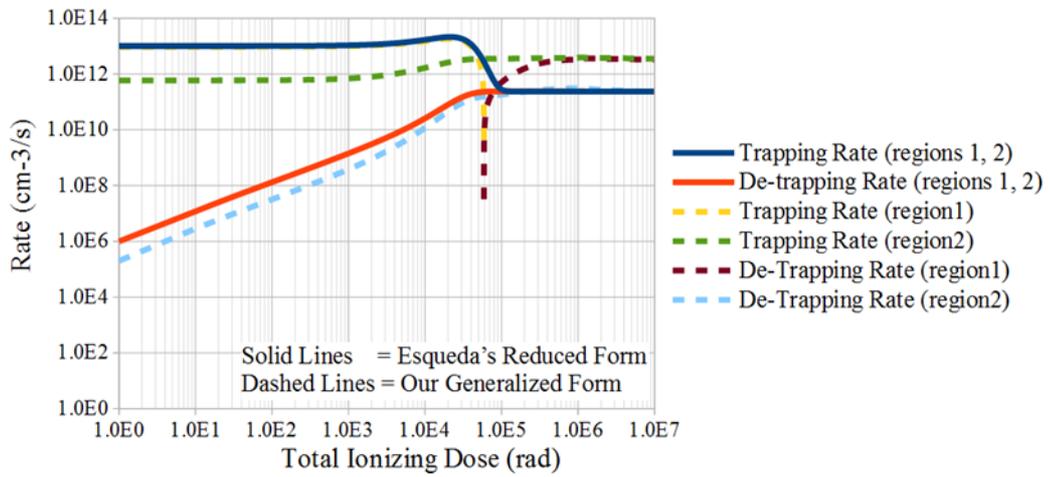

Figure 10: The trapping and de-trapping rates resulted from Esqueda's reduced model equations and from our generalized model equations (trapping rate from region1 stands for the trapping rate resulted from the generated holes in region 1, while de-trapping rate from region1 stands for the de-trapping rate resulted from the generated electrons in region 1) ($T_{ox}$=425nm, $V_{gb}$=1.0V)

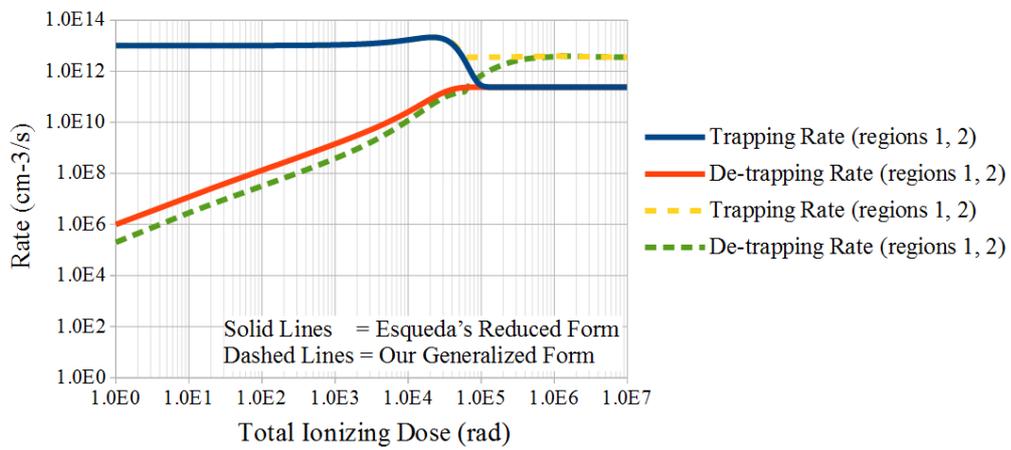

Figure 11: The net trapping and de-trapping rates resulted from Esqueda's reduced model equations and from our generalized model equations (Tox=425nm, Vgb=1.0V)

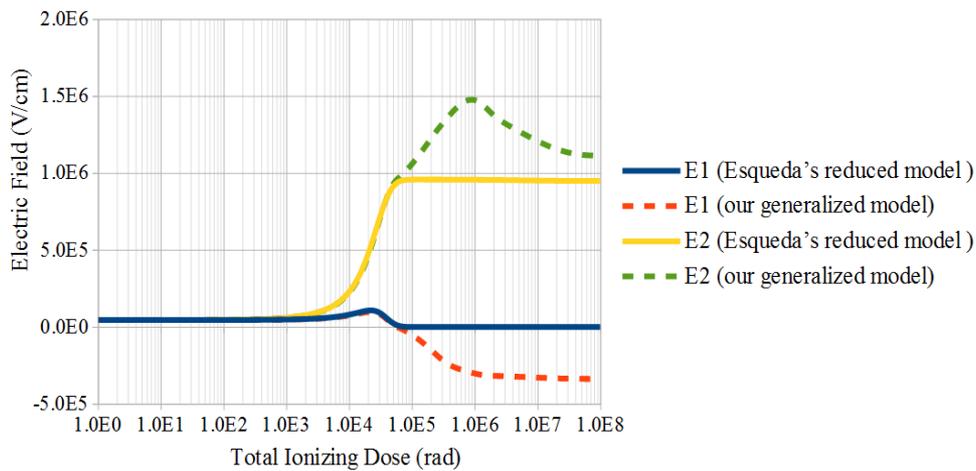

Figure 12: E1 and E2 as calculated using the Esqueda's reduced model equations and from our generalized model equations

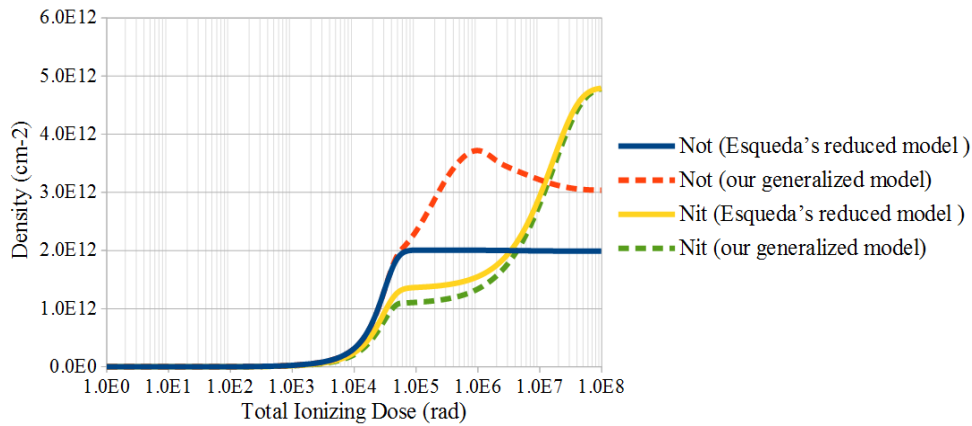

Figure 13: The oxide traps density and the interface traps density resulted from the Esqueda's reduced model equations and from our generalized model equations

## CONCLUSIONS

Accurate modeling of radiation effects on current MOSFET devices is critical to design of radiation-hardened ICs. We introduce improved models for the buildup of oxide traps and interface states in STI oxides of MOSFETS due to ionizing radiation exposure. The introduced models account for trapping and interface state buildup mechanisms for low as well as for high radiation exposure doses. They are also accurate for small as well as thick oxides. The introduced models fit published experimental results better than the currently published models.